\documentstyle[11pt,aaspp4]{article}
\input psfig 
\def\bibi{\bibitem}
\def\Mesz{M\'esz\'aros}
\def\mathnew{\mathsurround=0pt}
\def\simov#1#2{\lower .5pt\vbox{\baselineskip0pt \lineskip-.5pt
        \ialign{$\mathnew#1\hfil##\hfil$\crcr#2\crcr\sim\crcr}}}
\def\simg{\mathrel{\mathpalette\simov >}}
\def\siml{\mathrel{\mathpalette\simov <}}

\slugcomment{to appear in Astrophys. J., (Letters)}

\begin{document}
\begin{center}

\title{
SPECTRA OF UNSTEADY WIND MODELS\\
OF GAMMA-RAY BURSTS
}
\author{H. Papathanassiou \altaffilmark{1}  and P. M\'esz\'aros\altaffilmark{2}}
\affil{525 Davey Laboratory, Pennsylvania State University, University Park,
 PA 16802}
\altaffiltext{1}{hara@astro.psu.edu}
\altaffiltext{2}{Center for Gravitational Physics and Geometry, 
Pennsylvania State University; nnp@astro.psu.edu}
\end{center}
\bigskip

\begin{abstract}
We calculate the spectra expected from unsteady relativistic wind models of 
gamma-ray bursts, suitable for events of arbitrary duration.
The spectral energy distribution of the burst is calculated over 
photon energies spanning from eV to TeV, for a range of event durations 
and variability timescales. The relative strength of the emission at 
different wavelengths can  
provide valuable information on the particle acceleration, radiation 
mechanisms and the possible types of models.

\end{abstract}


\section{ Introduction}
\label{Introduction}

Relativistic wind models of Gamma Ray Bursts (GRB) are a recent development
of the dissipative relativistic fireball scenario which  is a natural 
consequence of  observationally set requirements  irrespective of  the
 GRB distance scale (e.g.~\cite{me95}). 
 The observed bursts are expected to be produced in optically 
thin shocks in the later stages of the fireball expansion. 
Two  types of shocks have been considered, based on different
interpretations of the burst duration and variability.  First,
the ``external'' deceleration shocks (\cite{rm92}; \cite{mr93}) 
that develop at $r_{dec}$, due to the unavoidable interaction of the relativistic ejecta with the
surrounding  medium, give bursts whose duration 
 is determined (in the ``impulsive'' regime) by the relativistic 
dynamic timescale $t_{dec} \sim r_{dec}/c \Gamma^2$, where $\Gamma$ is
the Lorentz factor of the expansion (e.g. \cite{mr93}; \cite{ka94};
Sari, Narayan \& Piran 1996). 
Second, the 
``internal'' dissipation shocks in an unsteady relativistic wind 
outflow (\cite{rm94} [RM94]; \cite{paxu94}), where both the burst duration
$t_w$ and time
variability  $t_{var} \leq  t_w$, are 
 characteristic of the progenitor mechanism (e.g. a disrupted disk accretion 
timescale, dynamic or turbulent timescales, etc.).  Simple ``external'' shocks
 tend to  produce fairly smooth light
curves with a fast rise followed by an exponential decay (FRED), as observed in  
some bursts. However, many  bursts have multi-peaked, irregular light curves; 
those may find a more natural explanation  (Fenimore, Madras \& Nayakshin 1996)
 in  the relativistic wind scenario mentioned above. 

While the spectral properties of impulsive ``external'' shocks have been studied 
in some detail (\cite{mrp94} [MRP94]; \cite{sa96}), so far, only rough estimates have 
 been made for wind ``internal'' shock spectra (\cite{mr94};  
\cite{th94}).
It is of great interest to explore the spectral properties of both types of shock scenarios,
particularly in view of the increasingly sophisticated analyses of both GRB 
light curves (e.g.  \cite{fe96}; \cite{nor96}) and spectral characteristics (e.g. 
\cite{band93}).
In this letter, we investigate the spectral properties of  ``internal''
shocks for representative  parameter values and discuss those properties in the light of
anticipated results from HETE. 

\section{  Physical Model and Shock Parameters}
\label{sec:model}

In the unsteady relativistic wind model, the GRB event is caused by 
the release of an amount of  energy $E=10^{51}E_{51}$ erg, inside a volume of 
typical dimension $r_{\ell}= 3 \times 10^{10} t_{var}$ cm, over a timescale 
$t_w$  (RM94). The average dimensionless entropy per particle $\eta= 
\langle E/(M c^2) \rangle - 1 = \langle L/({\dot M} c^2)\rangle$ determines 
the character of the overall flow. The bulk Lorentz factor of the flow 
increases as $\Gamma \propto r$ at first  and later saturates to a value 
$\Gamma = \Gamma_s$ at $r \simg r_s \approx r_{\ell} \Gamma_s $,
where $\Gamma_s \sim \eta $ for the values of $\eta$ that we consider
here (eq. [\ref{eta_lim}]). It is assumed that $\eta$ varies substantially,
$\Delta \eta \sim \eta$, on a timescale $t_{var} $.  As an example,
we consider shells of matter with two different  $\eta$ values  ejected at 
time intervals $t_{var}$. After saturation these will coast with bulk Lorentz 
factors $\Gamma_f \sim q_f \Gamma_s $ and $\Gamma_r \sim q_r \Gamma_s$,  
where $  q_r < q_f \sim 1$. Shells of different $\eta$ catch up with each other 
at a dissipation radius 
$r_{d} \approx \Gamma_{f} \Gamma_{r} c t_{var} \approx 3 \times 10^{14}\;
 q^2 \Gamma_{s_2}^{2}  \; t_{var} \; $ cm (see RM94, eq. [4]),
where $q =\sqrt{q_f q_r}$ and the subscript 2 (3) stands for quantities
measured in  units of $10^{2}$ ($10^3$). 
This gives rise to a $\em{forward}$ and a $\em{reverse}$ shock which
are mirror images of each other in the contact discontinuity (center of 
momentum, CoM) frame. The CoM  moves in the lab frame with 
$\Gamma_{b}\approx \sqrt{ \Gamma_{f} \Gamma_{r}}$, for $\Gamma_s \gg 1$. 
In the CoM frame the two shocks move in opposite directions 
with small Lorentz factors $\Gamma'_{sh} -1  \approx (1/2)
\;[({\Gamma_{f}}/{\Gamma_{r}}) + ({\Gamma_{r}}/{\Gamma_{f}})]^{1/2}-1
\ge 10^{-2}$, for $\Gamma_{f}, \Gamma_{r} \gg 1$, while in the lab frame both  
shocks move forward with Lorentz factors $\Gamma_{f}$ and $ \Gamma_{r} $ 
respectively. The shocks compress the gas to a comoving frame 
( primed quantities refer to the $\em{comoving}$ frame, which is the  CoM frame )
density
\begin{equation}
 n\acute{}\; (r_{d}) \approx \frac{8 \times 10^{9}}{q^4}
 \frac{E_{51}}{\theta ^{2} t_{w} t_{var}^{2}} 
\frac{1} { \eta_{2} \Gamma_{s_2}^{5} } \; \; \; \rm{cm^{-3}},
\end{equation}
where $\theta^{2}$ is the normalized opening solid angle of the wind ($1$ if 
spherical). The bulk kinetic energy carried by the flow is randomized in the 
shocks with a mechanical efficiency 
  $\varepsilon_{sh} \approx (q_f + q_r - 2 q)/ (q_f+ q_r)$ (eq. [1] in RM94).
In order for the shocks to produce a non-thermal spectrum and radiate a 
substantial portion of the wind energy, the dissipation must occur beyond both 
the wind photosphere (see section 2.2 in RM94) and the saturation radius. This restricts $\eta$ to the 
range 
\begin{equation}
 33 \; \left( \frac{E_{51}}{\theta^{2} t_{w} t_{var}} \right)^{1/5} \siml \eta 
\siml 80 \; \left( \frac{E_{51}}{\theta^{2} t_w t_{var}}\right)^{1/4}.
\label{eta_lim}
\end{equation}

The collisionless ``internal'' shocks 
accelerate particles to relativistic energies with a nonthermal distribution. 
 We 
parameterize the post-shock electron energy by a factor $\kappa$, so that the 
average 
random electron Lorentz factor is $\gamma_e \sim \kappa\Gamma\acute{}_{sh} \sim
\kappa \le m_p/m_e $ (with the  protons' being $\gamma_p \sim 
\Gamma\acute{}_{sh}$). An injection fraction $\zeta$ of the post-shock electrons 
is accelerated to a power-law
distribution ($ d N(\gamma) \sim \gamma^{-p} d\gamma$, for $\gamma_{min} 
\leq \gamma $). 
We take 
 $p \ge 3$, so that most of the energy is carried by the 
low energy part of the electron spectrum ($\gamma_{min} \approx ((p-2)/(p-1)) 
\kappa$). A measure of the efficiency of the transfer of energy between protons 
and electrons behind the shock is given by $\varepsilon_{pe} \simeq 
\zeta \kappa/(m_p/m_e) \sim \zeta\kappa_3$. Here, 
 we will use $\zeta=1$ and $\kappa\le 10^3$
( but see  \cite{byme96}) and $p=3$ which is consistent with the fits
to observed spectra (\cite{hanlon94}).
 The electrons get 
accelerated at the expense of the randomized proton energy behind the shocks, 
which, for an individual shock (of which there are $t_w/t_{var}$), represents a 
fraction $ ({t_{var}}/{t_w}) \varepsilon_{pe} \varepsilon_{sh} $ of the total 
energy $E_0$. 

Magnetic fields in the shocks can be due to a frozen-in component from the 
progenitor, or  may build up by turbulence behind the shocks. 
We parameterize the strength of the magnetic field in  the shocks
by the fraction $\lambda$ of the magnetic energy density to the particle 
random energy density ($u\acute{}_{\scriptscriptstyle B} = \lambda
n\acute{} m_p c^2$). The comoving magnetic field therefore is 
\begin{equation}
  B\acute{} \approx
\frac{1.72\times 10^{4}}{ q^2 \Gamma_{s_2}^2}   \;
\sqrt{\frac{ E_{51}\varepsilon_{sh}}
{\theta^{2} t_w t_{var}^{2}} \;
\frac{\lambda} {\eta_2 \Gamma_{s_2}}} \quad \hbox{G}.
\end{equation}

The relativistic electrons will lose energy due to synchrotron radiation,
and inverse Compton (IC) scattering of this radiation. The respective radiative efficiencies are
$\varepsilon_{sy} = t_{sy}^{'-1}/(t_{sy}^{'-1}
+t_{ic}^{'-1} +t_{ex}^{'-1})$, and 
$\varepsilon_{ic} = t_{ic}^{'-1}/(t_{sy}^{'-1}
+t_{ic}^{'-1} +t_{ex}^{'-1}) $ . The timescales are defined below.
The comoving synchrotron timescale is determined by the least energetic 
electrons : 
\begin{equation}
t'_{sy} \approx 8 \times 10^8 /\gamma_{min} B^{'2} =
5.2 \; \displaystyle  q^4 \;
\frac{\theta^2 t_w t_{var}^2}{E_{51} \varepsilon_{sh}}\;
\frac{\eta_2 \Gamma_{s_2}^5}{\lambda \kappa_{sy_3}}
 \; \; \; \hbox{ms},
\end{equation}
where the subscript {\em sy} ({\em ic}) refers to the synchrotron (IC) emitting
shock.
The IC cooling depends on, and competes with, synchrotron cooling.
IC cooling dominates if the magnetic field is weak, a large fraction
of electrons are accelerated and share  the protons' momentum  very efficiently 
(i.e. $\lambda  /(\zeta \kappa_{ic_3} )\ll 1$);
while synchrotron cooling dominates in the opposite case. 
The IC timescale ($t'_{ic} \approx 3 \times
10^7/u\acute{}_{sy} \gamma_{min}$) for the two limiting cases is 

\begin{equation}
t'_{ic} \approx 
\left\{ \begin{array}{ll} 
9.6  \;\displaystyle \frac{q^4}{\zeta \varepsilon_{sh}}
 \frac{\theta^2 t_w t_{var}^2}{E_{51}}
 \frac{\eta_2 \Gamma_{s_2}^5}{<\gamma_3^2>}
\frac{\gamma_{*_3}}{\kappa_{ic_3}}
\left[1 +\frac{t\acute{}_{sy}(\gamma_*)}{t\acute{}_{ex}}\right]
& \mbox{ms $\;\;\;$if IC dominates}\\ 
\label{t_ic_ic}
\\
4.95\;  \displaystyle \;\frac{q^4}{\varepsilon_{sh}}
 \frac{\theta^2 t_w t_{var}^2}{E_{51}}\;
\frac{\eta_2 \Gamma_{s_2}^5}
{\sqrt{\zeta \lambda \kappa_{ic_3} <\gamma_3^2>}}
& \mbox{ms \quad if synchr. dominates}
 \label{eq:t_ic_sy}.
\end{array}
\right.
\end{equation}
Here $\gamma_* $ is the electron Lorenz factor  that corresponds
to the peak emission frequency, 
$\gamma_* =max \left[\gamma_{min}, \gamma_{abs} \right]$,
 $\gamma_{abs}$ is defined below equation (\ref{v_abs}) and 
$ <\gamma^2_3> = 2\times 10^{-6}
\displaystyle \gamma_{min}^2
\;ln(\gamma_{max}/{\gamma_*})$.

The lab frame shell width (shocked region) is $\Delta r \sim r_d \Gamma_{b}^{-2}
\sim \alpha c t_{var}$, and the comoving crossing time 
$t_{ex}' \approx 10^{2} (\alpha q^2) \; \Gamma_{s_2} \;t_{var}$ s
provides an estimate for the adiabatic loss time.

The spectrum of an ``internal'' shock burst consists of two
synchrotron and four IC  components (coming from all shocks' combinations).
The synchrotron components are characterized by up to three {\it break 
frequencies}, given below in the lab frame:

i) The $\gamma_{min}$ break frequency $\nu' = 8 \times 10^5 B' \gamma_{min}^{2}$, which in the lab 
frame gets blue-shifted by $\Gamma_{f}$ ($\Gamma_r$) for the forward
(reverse) shock. Using $q_{sh}=[q_f,q_r]$ to refer to the shock, we have

\begin{equation} 
h\nu_{min} \approx 1.45\; \displaystyle 
\;\frac{q_{sh}}{q^2} \; \frac{\kappa_{sy_3}^2}{ \Gamma_{s_2} }
\;\sqrt{\frac{E_{51} \varepsilon_{sh}}{\theta^2 t_w t_{var}^2}
\frac{\lambda}{\eta_2 \Gamma_{s_2}}} \quad \hbox{keV}.
\label{v_sy_min}
\end{equation}

ii) The self absorption frequency $\nu_{abs}$. If the electron power law 
starts at low enough energies, the radiation field becomes optically thick 
at a frequency determined by $\frac{3}{2} m_{e} \gamma_{abs} \nu_{abs}^{'2} = 
F'_{\nu_{abs}}$, and is obtained by solving a non-linear algebraic
equation; in the limit $\nu_{abs} \gg \nu_{min}$ 
it is 
\begin{equation}
h \nu_{abs} \approx  18.4  \;\displaystyle \frac{q_{sh}}{q^{10/7}}\;
\frac{1}{\Gamma_{s_2}}  \left( \frac{\lambda}{\eta_2 \Gamma_{s_2}} \right)^{1/14}
\left ( \frac{E_{51}\varepsilon_{sh} } 
{\theta^{2} t_{w} t_{var}^{2}} \right)^{5/14}\;
(\zeta \varepsilon_{sy}  \kappa_{sy_3})^{2/7}
 \mbox{ eV $\; \; \;$ for $\gamma_{min} < \gamma_{abs}$},
\label{v_abs}
\end{equation}
where $\gamma_{abs} = 9 \times 10^3 \sqrt{\nu_{abs}/(q_{sh}
\Gamma_{s_2} B\acute{})}$.

iii) The frequency where the photon spectrum of the minimum energy electrons
becomes optically thick ($\nu_{\scriptscriptstyle{RJ}}$), and below
which it assumes a  
Rayleigh-Jeans spectrum slope. This happens at
$\nu_{min}$ when $\gamma_{min}< \gamma_{abs}$; if $\gamma_{min} \gg
\gamma_{abs}$ it is 
\begin{equation}
h\nu_{\scriptscriptstyle{RJ}} \approx 
0.15 \; \displaystyle \frac{q_{sh}}{q^{4/5}}
\left(\frac{E_{51}    \varepsilon_{sh}}
{\theta^{2} t_{w} t_{var}^{2}} \right)^{1/5}
\left(\frac{\eta_2}{\lambda} \right)^{2/5} 
\left(\frac{\zeta \varepsilon_{sh}}{\Gamma_{s_2}}\right)^{3/5}
\frac{1}{\kappa_{sy_3}^{8/5}} \; \; \hbox{eV}.
\label{eq:v_RJ}
\end{equation}

iv) A  cutoff is expected at 
$\nu_{max} =
(\gamma_{max}/\gamma_{min})^{2} \; \nu_{min}$,
 where $\gamma_{max}$ is the electron energy in each shock at which the electron 
power law cuts off due to radiative or adiabatic losses. It is  determined by 
$t_{acc}(\gamma_{max}) \approx 
t'_{cool}(\gamma_{max})$, where $t_{acc}$ is the electron acceleration timescale in 
the shocks (a multiple $10 \times A_{10}$ of the inverse gyro-synchrotron frequency), 
$t_{acc} \approx {3.57 \times 10^{-6} A_{10}}{B'(r_{d})^{-1}} \gamma$ s,
and $t_{cool}$ is the minimum of all the radiation and adiabatic cooling 
timescales involved,
$t'_{cool_{r,f}} = min \left\{ t_{ex}', t'_{sy_{f,r}}, t'_{IC_{r}, } 
t'_{IC_{f}}  \right \}$ .

Each synchrotron component can be characterized by three frequencies in ascending 
order, $\nu_{sy,j}^{i}$, where $i=1,..,3$ and $j=1$ for the reverse  and 
$j=2$ for the forward shock. Similarly, the pure and combined IC spectra are
characterized by the frequencies $\nu_{ic,j}^{i} \approx ({4}/{3})  
\;\kappa_{ic}^{2} \nu_{sy,l}^{i}$ where, $j=1,..,4$ 
(1 corresponds to {\it IC reverse} , 2  to {\it IC forward}, 3 to {\it IC 
reverse-forward} and 4 to {\it IC forward-reverse}), and if j is odd,
$l=1$, otherwise $l=2$. The shape of each component depends on the 
relationship between the relevant $\gamma_{min}, \gamma_{abs} $ and $\gamma_{max}$.
In a power per logarithmic frequency interval plot, the fluence of
each component exhibits a peak of
$ S_{i} = 1.6 \times 10^{-6} (E_{51}/(\theta D_{28})^{2})  \;
\varepsilon_{sh} 
\varepsilon_{i} \zeta \kappa_{i_3} \; \; {\rm erg/cm}^{2}$, 
where $i=1 (2)$ for synchrotron (IC), and $ D_{28}$ is the luminosity 
distance corresponding to $z \approx 1$ for a flat Universe, with $H_{o} \approx 
80$ ($D_{L} = (2 c/H_{o}) (1+ z - \sqrt{1+z} \approx 3 \; \hbox{Gpc})$). 
The spectrum is obtained by adding up these six components. In practice, the
forward and reverse components have values very close to each other and they 
essentially merge. The resultant spectrum is then checked for  the effects of pair 
production; the $\gamma\gamma$ optical depth is calculated for 
each comoving frequency above $m_e c^2 /k$ using the number density of photons 
above the corresponding threshold, as obtained from the initial spectrum; 
finally the spectrum is modified accordingly. 
We note that most of the scattering in our spectra occurs in the Thomson regime,
unlike in ~\cite{sa96}, who consider large $\gamma$ in the framework of 
impulsive shock models. ( Klein-Nishina corrections may become relevant 
in a few $\kappa \siml 10^2$ cases, but at frequencies 
which lie in the $\gamma\gamma$ absorbed part of the spectrum). 
 
\section{ Typical Wind Spectra}

We discuss here the properties of some representative  spectra.
We assume a total  event energy of $E=10^{51}$ erg and a geometry of a 
spherical section, or jet, of opening angle $\theta = 0.1$ (the physics is 
the same as in a spherical wind , provided $\theta > \Gamma^{-1}$). 
We have investigated a range of dynamic parameters ($\eta, t_w$ and
$t_{var}$), and
  used  $q_f=2, q_r=0.5$, and $\kappa_f =\kappa_r$.

In figure {\ref{fig_long}} we present spectra for a long burst ($t_w =100 $
s) and
in figure {\ref{fig_short}} for a short one ($t_w= 1$ s), for different
$\eta, \lambda$ and $\kappa$ 
 values. 
The sharper spectral features present would be smoothed  by inhomogeneities in a real flow. 

The range of $\eta$ where a substantial fraction of the wind energy is
radiated with nonthermal spectra is considerably restricted 
(eq. [\ref{eta_lim}]) and implies lower values than in impulsive models.
Larger $\eta$ may be difficult to produce in a natural way, while
lower values lead to shocks below the photosphere that make bursts too
dim to be observed. The allowed  volume of parameter space is not very large
(\cite{ha96} [PM96]). Nonetheless, it allows for an appreciable variety of spectra, 
since models differing only slightly in $\eta$  can produce significantly
different spectra. This is   due to the strong
dependence of the photon number density on $\Gamma_s$
(i.e. $ n\acute{}_{\gamma} \propto \Gamma_s^{-7} $) which determines pair
production. Lower $\kappa$ spectra are less affected by pair production, 
since they contain less energetic electrons and therefore fewer photons.  
Generally, other parameters being equal, the higher $\eta$ models tend 
to produce spectra that span over a wider range of frequencies.
For spectra like the majority of the ones observed by BATSE (\cite{Fi&Me_rev})
values of $\kappa < 10^2$ and $\lambda > 10^{-1}$ are excluded. Low
$\kappa$ spectra are either too dim or too soft; high $\lambda$ brings up the
synchrotron component and may be appropriate for a small percentage of
bursts
with low frequency excess (\cite{Preece96}).
In general, for a given $\kappa$ value a wide range of $\lambda$ values (3 - 6
orders of magnitude) is allowed, the trend being that higher $\kappa$ values 
must combine with lower $\lambda$ values, fairly independently  of $\eta$ 
(within the allowed range of values). This trend is due to the fact that 
lower magnetic fields require higher electron energies in order for the break 
to fall in the BATSE window (e.g. eq. [\ref{v_sy_min}], [\ref{v_abs}]
). High energy power laws (like those reported in {\cite{hanlon94}})
are common (see fig. \ref{fig_short} and right column of fig. \ref{fig_long}).
For a more complete discussion see PM96.

The effect of a longer $t_w$ is to push the pair cutoff to higher
frequencies, because, for fixed $E_0$, the flux and photon density are lower.
For the same $\eta$, $\kappa$ and $t_{var}$, longer bursts require
higher $\lambda$. 
 A greater  $t_{var}$ with  the rest of the parameters unchanged would
again require  higher $\lambda$ in order to produce observed-like spectra.
The cases considered here were chosen
with $ t_w < t_{dec} \approx (E_0/n_{ex} m_p c^5 \eta^8)^{1/3}$ s.

\section{Conclusions}
\label{conclusions}
An unsteady relativistic wind provides an attractive scenario for the
generation of GRB, since it requires smaller bulk Lorentz factors than the
impulsive models, i.e. it can accommodate higher baryon loads (RM94).
In addition, the lack of kinematic restrictions (e.g. \cite{fe96}) allows it, 
in principle, to describe events with arbitrarily complex light curves. 
We have calculated spectra from  optical through TeV frequencies for bursts
with a range of durations and variability timescales.
At cosmological distances the total energies and photon densities implied by 
the model are likely to turn those spectra optically thick to 
$\gamma\gamma$ pair production. Most of them may therefore 
 be missed by, or show a high energy cut-off in, the EGRET window, but
they would be prominent in the BATSE and HETE gamma-ray windows.
Low frequency (down to 5 keV) excess reported recently
({\cite{Preece96}}) may be attributed to a pronounced synchrotron
component due to a relatively high magnetic field, in a fraction of
the bursts.

The continued propagation of an unsteady wind flow should eventually lead to its 
deceleration by the surrounding medium. If the latter is of any appreciable
density, it could lead to another burst with the ``external'' shock 
characteristics (MRP94, \cite{mr94}), provided that not all of the wind energy 
was radiated away by the ``internal'' ones ($\varepsilon_{sh} < 1$).
The spectra of the ``internal'' shocks are different from those coming
from the ``external'' ones, the main differences being that the former
cover a narrower range of frequencies and most have high energy cutoffs due to 
pair production opacity (PM96). If GRB progenitors have escaped
their parent galaxies and are in a low density intergalactic medium
($n_{ex} <10^{-3}$ cm$^{-3}$), the ``external'' impulsive shocks would be of long 
duration ($> 3\times10^3 /\eta_2^{8/3} $ s)  and low 
intensity, and most would be totally  missed, hence the GRB could be entirely 
due to unsteady wind ``internal'' shocks. If the GRB occur in a denser medium 
(e.g. the galactic ISM), the last peak of multiple-peaked GRB would be smooth 
and FRED-like but the earlier peaks, due to unsteady wind ``internal'' shocks, 
may be arbitrarily complicated.
Those might also be responsible for delayed GeV emission ({\cite{mr94}}).

\begin{figure}[htbp]
\centerline{\psfig{figure=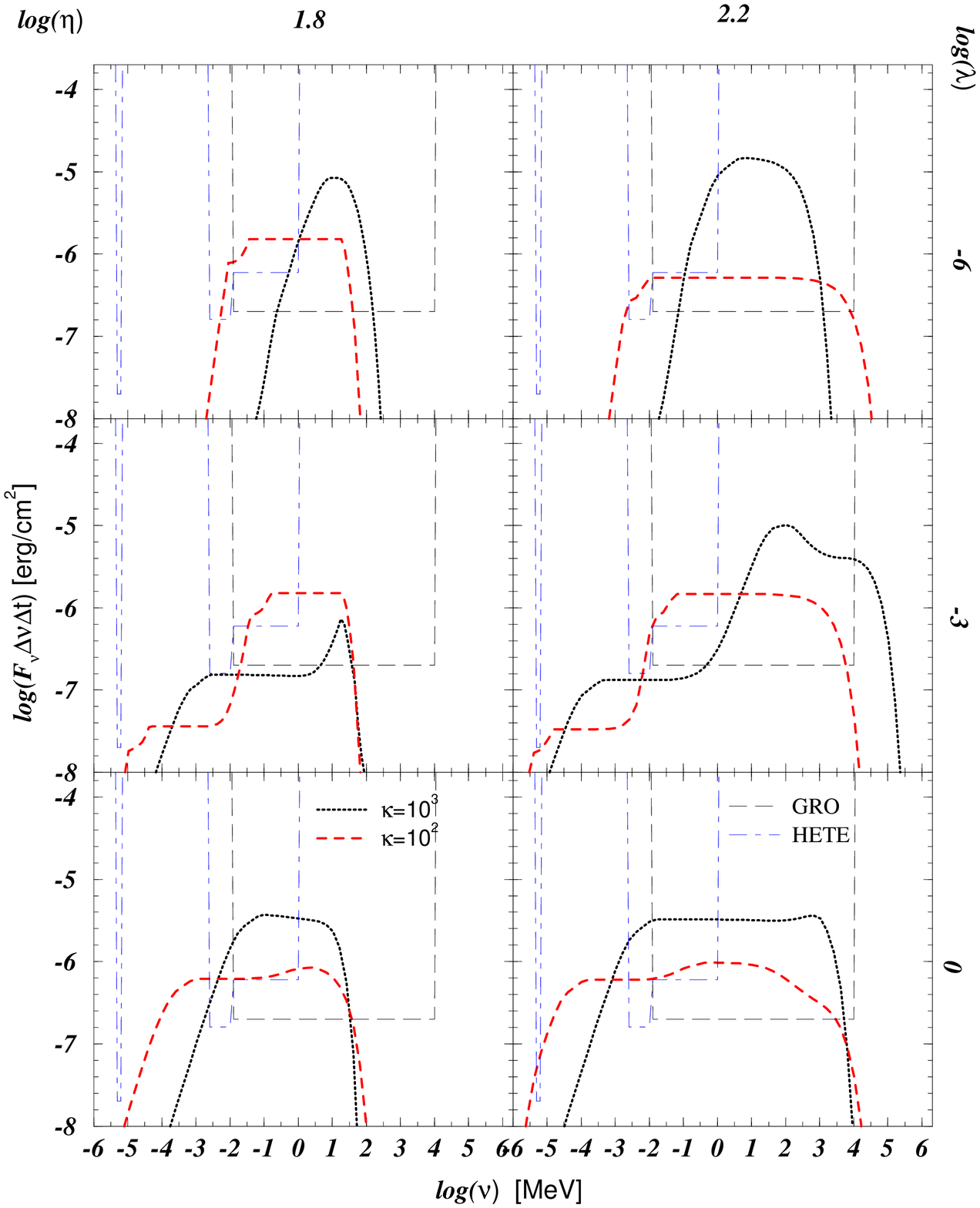,height=17.0cm,width=15.5cm}}
\caption{Spectra for $E_{51}=1,\theta =0.1, t_w=100 \rm{s}, t_{var}=40
\rm{ms}$, $\eta= 63$ (left column), $\eta\approx 160$ ( right
column), and for
different values of the magnetic field and the electron acceleration 
efficiency.
The long dashed line shows the approximate threshold and window for
GRO's BATSE and EGRET experiments, while the dotted-dashed one is for 
all the experiments on board HETE.}
\label{fig_long}
\end{figure}

\begin{figure}[htbp]
\centerline{\psfig{figure=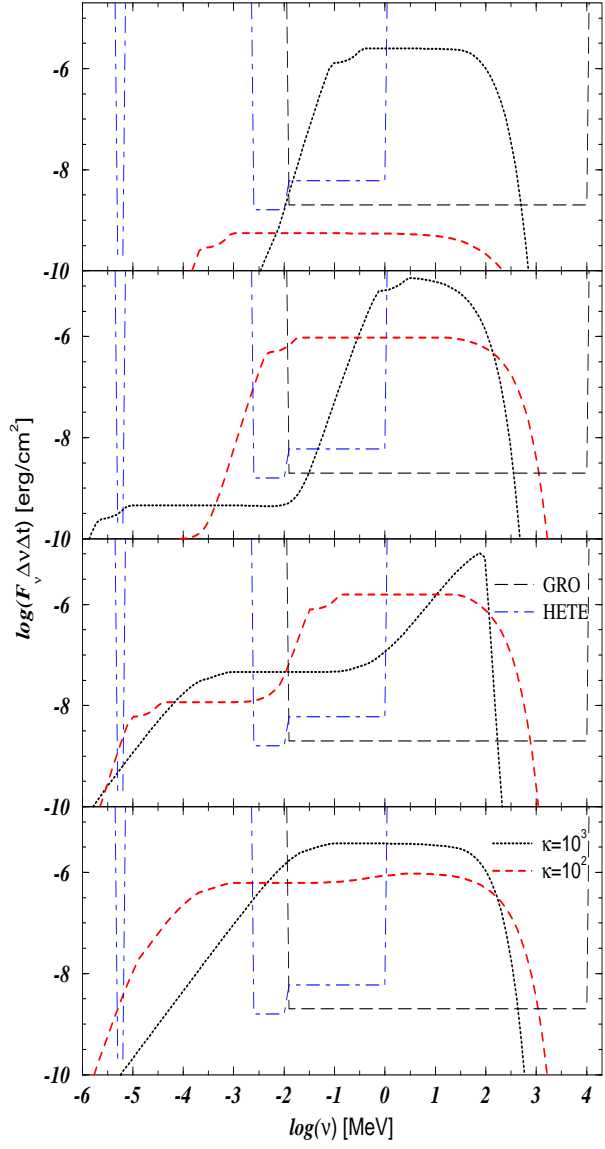,height=17.0cm,width=10.5cm}}
\caption{ Spectra for $E_{51}=1,\theta =0.1, t_w =1
\rm{s}, t_{var}=40 \rm{ms}, \eta= 200$ and for representative values of the
magnetic field (from bottom to top $log(\lambda)= 0,-4,-8,-12$)
and electron acceleration efficiency. GRO and HETE windows are included.}
\label{fig_short}
\end{figure}

A number of unknown parameters enter into the calculation of GRB models, and
HETE may provide the information required to narrow the range of their allowed 
values, as well as a test for the general scenario of unsteady relativistic 
winds. For the HETE sensitivities indicated in figures \ref{fig_long} 
and \ref{fig_short}, we expect that some bursts will show a simultaneous  
X-ray counterpart. However, detection by the Ultraviolet Transient Camera  
(UTC) on HETE should be rare for the majority of bursts (note though that 
figures \ref{fig_long} and \ref{fig_short}  refer to $z\sim$ 1
distances, and that at a few $\times 10^2$ Mpc a bright burst would be $30-100$ 
times brighter, thus increasing the likelihood of detection by the UTC). For
the majority of faint bursts, a detection in the ultraviolet would be
possible only for low values of $\kappa$, resulting in a very broad and flat 
spectrum with upper cutoffs, if any, only at the highest energies.

\acknowledgements{This research has been supported through NASA NAG5-2362, 
NAG5-2857 and NSF PHY94-07194. We are grateful to the Institute for Theoretical
Physics, UCSB, for its hospitality, and to participants in the Nonthermal 
Gamma-Ray Source Workshop for discussions}.

\pagebreak

\pagebreak

\end{document}